\documentstyle[prb,aps,multicol,psfig]{revtex}
\begin{document}
\draft

\title{Anomalous corrections to the Kelvin equation at complete wetting}
\author{Enrico Carlon{$^\dagger$}, Andrzej Drzewi\'nski{$^\ddagger$}
and Jos Rogiers{$^\dagger$}}
\address{{$^\dagger$} Institute for Theoretical Physics, Katholieke Universiteit Leuven,
Celestijnenlaan 200D, B-3001 Leuven, Belgium}
\address{{$^\ddagger$} Institute of Low Temperature and Structure Research, Polish
Academy of Sciences, P. O. Box 1410, 50-950 Wroc\l aw 2, Poland}
\date{published in \prb {\bf 58}, 5070 (1998)}

\maketitle

\begin{abstract}
We consider an Ising model confined in an $L \times \infty$ geometry with identical 
surface fields at the boundaries. According to the Kelvin equation the bulk coexistence 
field scales as $1/L$ for large $L$; thermodynamics and scaling arguments predict higher
order corrections of the type $1/L^2$ and $1/L^{5/3}$ at partial and complete wetting
respectively.
Our numerical results, obtained by density-matrix renormalization techniques for systems
of widths up to $L = 144$, are in agreement with a $1/L^2$ correction in the partial wet 
regime. However at complete wetting we find a large range of surface fields and temperatures
with a correction to scaling of type $1/L^{4/3}$. We show that this term is generated by a 
{\it thin} wetting layer whose free energy is dominated by the contacts with the wall.
For $L$ sufficiently large we expect a crossover to a $1/L^{5/3}$ correction as predicted
by the theory for a {\it thick} wetting layer. This crossover is indeed found in a 
solid-on-solid model which provides a simplified description of the wetting layer and 
allows the study of much larger systems than those available in a density-matrix 
renormalization calculation.
\end{abstract}


\begin{multicols}{2}\narrowtext

\section{introduction}
\label{sec:intr}

Wetting phenomena are very common in nature \cite{dietrich}; besides the most
familiar situation of a liquid-vapor or of a binary fluid mixture in contact 
with a solid wall, wetting has also been studied, for instance, at grain 
boundaries \cite{grain}, in polymer mixtures \cite{polymers}, alloys \cite{alloys},
superconductors \cite{SC} and metallic vapors \cite{metallic}.
Usually one considers a semi-infinite geometry with a solid planar wall that 
preferentially adsorbs one of the phases of a system in thermodynamic equilibrium. 
Below $T_c$, the bulk critical temperature, the adsorbed phase forms either isolated 
droplets or a thick macroscopic layer. The first case, known as {\it partial} wetting,
occurs for temperatures below a temperature $T_w$, the wetting temperature, while the 
second case occurs for $T_w \leq T < T_c$ and it is referred to as {\it complete} wetting.

In this paper we analyze the effect of wetting on the thermodynamics of a two-dimensional
Ising model confined in an $L \times \infty$ geometry, with identical surface fields at the 
boundaries. If $L \to \infty$, i.e. in the bulk, phase coexistence occurs for temperatures 
$T < T_c$ and for vanishing bulk magnetic field $h = 0$. It is well known that the combined 
effect of surface fields and confinement shifts phase coexistence to a finite value of the 
bulk magnetic field $h = h_0 (L) \neq 0$, which for large $L$ scales as:
\begin{eqnarray}
h_0 (L) = \frac 1 L \,\, \frac{\sigma_0 \cos \theta}{m_b} ,
\label{kelvin}
\end{eqnarray}
where $\sigma_0$, $m_b$ and $\theta$ are the surface tension, the bulk spontaneous 
magnetization and the contact angle, respectively. Equation (\ref{kelvin}) is known 
in the literature as the Kelvin equation and has a long history that dates back to the 
last century \cite{thomson}; it essentially states that to have equilibrium 
between two phases, or phase coexistence, one needs a bulk field of the order of $1/L$ 
that compensates the effect of the surface fields.

Albano {\it et al.} \cite{albano} and Parry and Evans \cite{andy} analyzed the next order 
correction term to the Kelvin equation. Both studies, using scaling and thermodynamics 
arguments, concluded that for temperatures below the wetting temperature $T_w$, i.e. 
at partial wetting, the leading correction to scaling term is of type $1/L^2$. In the case of 
complete wetting the correction is expected to be non-analytic due to a singularity of 
the surface free energy. For the two-dimensional Ising model \cite{albano,andy} the 
predicted correction term is proportional to $1/L^{5/3}$.
Monte Carlo simulations on $M \times L$ lattices with $M \gg L$ were also 
performed \cite{albano}; the dominant $1/L$ scaling of the Kelvin equation was well 
verified (this was done also in three dimensions \cite{binder}), but the data were 
not accurate enough to convincingly test the type of corrections to scaling.

We have studied this problem using density-matrix renormalization group \cite{white} 
(DMRG) techniques and we have calculated the bulk coexistence field $h_0(L)$ for several 
values of surface fields and temperatures for strips of widths up to $L = 144$. Our 
numerical data are in good agreement with corrections of type $1/L^2$ in the partially wet 
regime, but not compatible with $1/L^{5/3}$ corrections at complete wetting.
We find instead that the correction to scaling at complete wetting is in good agreement,
for a wide range of surface fields and temperatures, with a term of type $1/L^{4/3}$. 
An explanation of this discrepancy is discussed in detail in the paper.

The DMRG has been applied mostly to the study of ground state properties of quantum 
systems both in one \cite{onedquantum} and two dimensions \cite{twodquantum}. The method was 
originally developed by White \cite{white} for the diagonalization of quantum Hamiltonians 
of very large systems using a truncated basis set. It does not have the minus sign 
problems that plague quantum Monte Carlo simulations of fermionic systems at low temperatures.

Nishino \cite{nishino} applied White's DMRG to two-dimensional classical systems, which 
are deeply related to quantum systems of one dimension less \cite{suzuki}. 
In the classical case a transfer matrix of a large strip is generated by a series of 
iterations that, starting from a small strip, enlarge the system and truncate the 
configurational space very efficiently. 
Although there are obviously no minus sign problems in the classical case the advantages 
of the DMRG with respect to the Monte Carlo method are mainly the high accuracy and the 
possibility of dealing with very large systems. Untill now the method has been applied
only exclusively to two-dimensional classical systems \cite{twodclassical}; although 
extensions to three dimensions are possible, they will require a high computational effort 
and one should not expect the same high accuracy that is typically found in two dimensions.

Recently a series of classical models in a confined geometry have been 
investigated using DMRG techniques; the studies focused on the effect of gravity on phase 
coexistence \cite{ourPRL} and on the critical point shift \cite{ourPRE} for a confined fluid, 
on the calculation of critical density profiles and Casimir amplitudes for Potts 
models \cite{ferenc} and on studies of critical adsorption \cite{ania}. 
Another advantage of the DMRG for these types of models is that, being based on a
transfer matrix approach, it naturally deals with geometries which are confined along 
one direction only and infinite in the other, i.e. with $L \times \infty$ lattices. 
More details on the DMRG can be found in the broad existing literature; a short review on 
the method has been presented in Ref. \onlinecite{review}.

The paper is organized as follows: In Sec. \ref{sec:mod} we introduce the model and 
present the numerical results focusing on the corrections to scaling for the Kelvin
equation; in Sec. \ref{sec:theory} we review the existing theory leading to a correction 
to scaling term proportional to $1/L^{5/3}$ for the two-dimensional Ising model. 
In Sec. \ref{sec:wettinglayer} we discuss a mechanism that explains the apparent 
discrepancy of the DMRG results with the theory. In Sec. \ref{sec:capill} we consider a 
discrete solid-on-solid model and analyze the correction to scaling exponent for that 
case. Sec. \ref{sec:concl} terminates our paper with some conclusions.

\section{The DMRG results}
\label{sec:mod}

We consider the Ising model on an $L \times \infty$ lattice with the following 
Hamiltonian:
\begin{eqnarray}
H = -J \sum_{\langle ij \rangle} s_i s_j - h_1 {\sum_k}^{({\rm I})} s_k - h_1 
{\sum_l}^{({\rm II})} s_l - h \sum_i s_i ,
\label{ham}
\end{eqnarray}
with $J >0$ and $s_i = \pm 1$.
The first sum is restricted to nearest neighbors, while (I) and (II) indicate 
the sums performed on spins at the boundary only; $h_1$ is a surface field 
which models the preferential adsorption of the boundaries for one of the two phases, 
and $h$ is a bulk magnetic field.
In the rest of the paper we take $J = 1$, which fixes the bulk critical temperature
to $T_c = 1/\ln (1 + \sqrt{2} ) \simeq 2.269$.

\begin{figure}[b]
\centerline{
\psfig{file=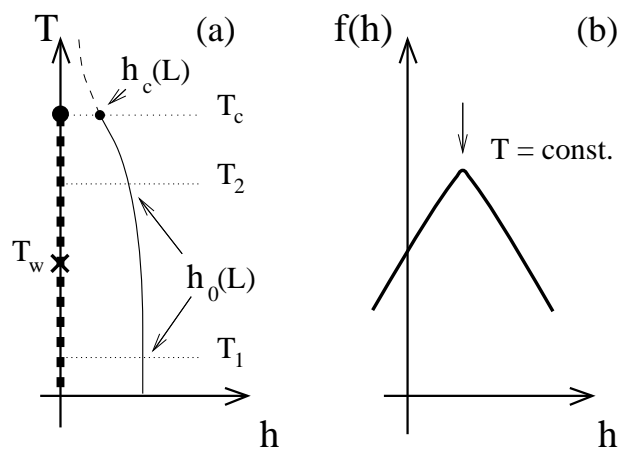,height=6cm}
}
\vskip 0.2truecm
\caption{
Due to the finiteness of $L$ and to the surface fields the {\it pseudo} phase-coexistence 
(thin solid line) is shifted with respect to the bulk coexistence line (thick dashed line). 
The temperatures $T_1$ and $T_2$ are in the partially and completely wet regimes respectively. 
(b) The pseudo-coexistence point (indicated by the arrow) is identified with the maximum 
of the free energy as function of the bulk field $h$.}
\label{FIG01}
\end{figure}

The phase diagram for the model for fixed $L$ and $h_1$ is shown in Fig.~\ref{FIG01}(a). 
The thick dashed line indicates the bulk phase coexistence line $h=0$, $T < T_c$ terminating 
at the bulk critical point. For finite $L$ there are no true thermodynamic singularities and 
quantities such as the free energy and the magnetization depend smoothly on temperature, 
surface and bulk fields.
Fig.~\ref{FIG01}(b) shows a plot of the total free energy $f(h)$ at fixed $T$, $L$, 
$h_1$: For large $h$, either positive or negative, the system is ``frozen" with all 
spins pointing up or down and the free energy as a function of $h$ is a straight line 
with slopes $\pm 1$.
In three or higher dimensions these two branches meet each other forming a cusp at 
phase coexistence or with higher order singularities at the finite system critical 
point. In two dimensions the singularities are replaced by a smooth maximum as 
indicated in Fig.~\ref{FIG01}(b). For $T < T_c$ we identify this maximum with the 
pseudo-coexistence field $h_0(L)$ which is expected to scale according to the Kelvin 
equation (\ref{kelvin}).

The free energy maximum at $T = T_c$ allows us to identify a critical field
$h_{\rm c}(L)$, whose scaling properties are governed by critical exponents.
Finite size scaling predicts \cite{nakanishi} that for large $L$:
\begin{eqnarray}
h_{\rm c} (L) \sim L^{-y_{\rm H}} ,
\label{scalmag}
\end{eqnarray}
with $y_{\rm H} = 15/8$ for the two-dimensional Ising model. 
Notice that even for $T > T_c$ one can find free energy maxima; the locus of these 
points is indicated by a thin dashed line in Fig.~\ref{FIG01}(a) and approaches
the $h=0$ axis exponentially in $L$. For our purposes this is an uninteresting region
and it has not been investigated.
It should be stressed that the definition of $h_{\rm c} (L)$ is somewhat arbitrary. In higher
dimensions one can find unambiguosly a finite system critical point and it is well
known that, due to the effect of confinement, both magnetic field and temperature 
are shifted \cite{nakanishi} with respect to the bulk critical point. Also in the present 
case it would be possible to define a finite system pseudo-critical point identified as 
the point for which, for instance, the specific heat has a maximum along the thin solid 
line of Fig.~\ref{FIG01}(a); however we focus here on the shift along the magnetic field 
direction and the definition given above of $h_{\rm c} (L)$ is appropriate and definitely less 
computationally demanding.

In the rest of this Section we will present the DMRG results concerning
the scaling analysis of the pseudo-critical ($h_{\rm c}(L)$) and the pseudo-coexistence 
($h_0(L)$) fields. We start from the first case, which will be just briefly discussed.

\subsection{$T = T_c$}

\begin{figure}[b]
\centerline{
\psfig{file=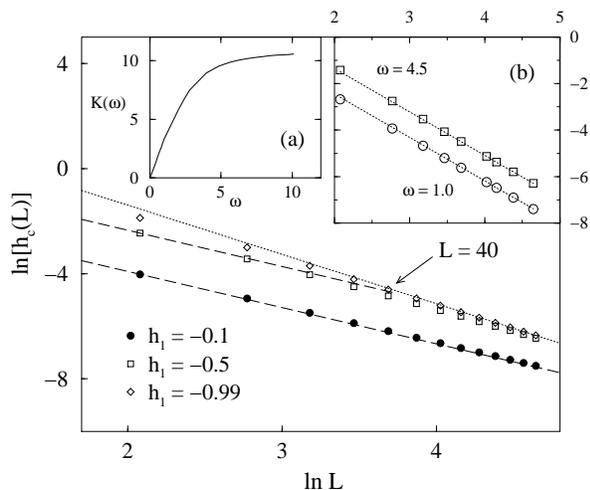,height=7cm}}
\vskip 0.2truecm
\caption{Crossover between an exponent $y_{\rm H} - \Delta_1 y_T = 11/8$ 
(the dashed lines) and an exponent $y_{\rm H} = 15/8$ (the dotted lines) 
for the scaling of the critical field $h_{\rm c}(L)$. 
Insets: (a) Plot of the scaling function $K( \omega)$ vs. $\omega$ calculated
from the numerical DMRG data. (b) Plots $\ln [h_{\rm c} (L)]$ vs. $\ln L$ for 
constant values of the scaling variable $\omega = h_1 L^{\Delta_1 y_T}$.
}
\label{FIG02}
\end{figure}

Figure~\ref{FIG02} shows a plot of $\ln [h_{\rm c}(L)]$ vs. $\ln L$ from $L = 8$ up to 
$L = 104$ and for three different values of the surface field $h_1 = -0.1$, $-0.5$ and 
$-0.99$.
The dotted line has a slope equal $-15/8$, which is the expected value of the exponent 
$- y_{\rm H}$ for the two-dimensional Ising model and is in good agreement with the 
data only for the highest surface fields considered ($h_1 = -0.5$, $-0.99$).

The behavior of $h_{\rm c}(L)$ for smaller surface fields can be understood from
finite-size scaling: The singular part of the surface free energy depends on the
surface field through the scaling variable \cite{barber} $h_1 L^{\Delta_1/\nu}$, 
where $\Delta_1$ is the surface gap exponent (recall $\Delta_1 = 1/2$ for the 
two-dimensional Ising model). Thus the surface field enters in the analysis as 
follows \cite{nakanishi}:
\begin{eqnarray}
h_{\rm c} (L) = L^{-y_{\rm H}} \,\,\,K \left( h_1 L^{\Delta_1/\nu} \right) ,
\label{scalingfunc}
\end{eqnarray}
where $K(\omega)$ is a scaling function.
For large values of its argument the scaling function $K(\omega)$ should 
``saturate" to a constant non-vanishing value $K_{\infty} \neq 0$, since one 
expects a scaling behavior as described in Eq. (\ref{scalmag}) 
for sufficiently large $L$. In absence of surface fields there is no shift
along the magnetic direction, i.e. $h_{\rm c} (L) = 0$, which implies $K(0) = 0$.
For small $\omega$ one expects $K(\omega) \sim \omega$ since $h_{\rm c} (L)$
should scale linearly with the surface field.
This leads to the following prediction:
\begin{eqnarray}
h_{\rm c} (L) \sim  L^{-\left( y_{\rm H} - \Delta_1/\nu \right)} ,
\label{crosshc}
\end{eqnarray}
for $h_1$ small and $L$ not too large.
This scaling behavior is clearly seen in the numerical DMRG data for $h_{\rm c} (L)$
shown in Fig.~\ref{FIG02} for the lowest surface field analyzed, i.e. $h_1 = -0.1$, 
where the dashed line has a slope equal $-11/8$ in agreement with the prediction of 
Eq.~(\ref{crosshc}).
From the data for $h_1 = -0.5$ one finds an estimate of the crossover size at
$L \approx 40$. 
The shape of the scaling function $K(\omega)$ found from the DMRG data is
shown in the inset (a) of Fig.~\ref{FIG02} and it is indeed linear for small 
$\omega$ and it is found to saturate to a constant $K_{\infty} \approx 10$ 
for $\omega > \omega_0 \approx 5$.

To verify that our numerical data fulfill a scaling of the type presented
in Eq. (\ref{scalingfunc}), we have plotted in the inset (b) of Fig.~\ref{FIG02}
again $\ln [h_{\rm c} (L)]$ vs. $\ln L$ for constant values of the scaling
variable $\omega = h_1 L^{\Delta_1/\nu}$. The data follow nice straight lines
with slope $-15/8$ even in the region of small $\omega$, i.e. for 
$\omega \ll \omega_0$.

Finally our estimate of the magnetic exponent $y_{\rm H}$ from the data of $h_{\rm c}(L)$
at $h_1 = - 0.99$ and $-0.5$ yield $y_{\rm H} = 1.874999(1)$ and $y_{\rm H} = 1.87506(1)$, 
respectively, in an excellent agreement with the exact value $15/8$.
Such a high numerical accuracy for critical exponents is typical of the DMRG
method used in combination with powerful extrapolation techniques \cite{ferenc}.

\subsection{$T < T_c$}

For temperatures below the bulk critical temperature $T_c \simeq 2.269$ one expects a scaling 
of the type (\ref{kelvin}). In practice however, since only a limited size $L$ is available 
for numerical computation, it is preferable to consider only temperatures not too close to 
$T_c$, where a crossover towards a scaling of the type of Eq.~(\ref{scalmag}) is expected. 
At low temperatures one also has the advantage that DMRG calculations become very rapid. 
In the present study we have calculated a series of values of $h_0(L)$ up to $L = 144$ 
and for temperatures ranging from $T \approx 0.97 \, T_c$ down to $T \approx 0.4 \, T_c$ 
with different values of the surface field $h_1$.

We introduce the logarithmic derivatives:
\begin{eqnarray}
x_L \equiv - \frac{ \ln \left[ h_0 (L) \right]
- \ln \left[ h_0 (L+8) \right]}{\ln  L-\ln (L+8)} ,
\label{defxL}
\end{eqnarray}
for $L =8$, $16$, $24$, \ldots, $136$. Assuming the following expansion:
\begin{eqnarray}
h_0 (L) = \frac{A}{L^\alpha} + \frac{B}{L^\gamma} + \ldots ,
\label{corrections}
\end{eqnarray}
where we expect $\alpha = 1$ and $A$ given by the Kelvin equation (\ref{kelvin}) and 
$\gamma > \alpha$ a correction-to-scaling exponent, one has to lowest orders in $1/L$:
\begin{eqnarray}
x_L = \alpha \left(1 - \frac{B}{A} \,\, \frac{1}{L^{\gamma - 
\alpha}} + \ldots \right) .
\label{complete}
\end{eqnarray}

\begin{figure}[b]
\centerline{
\psfig{file=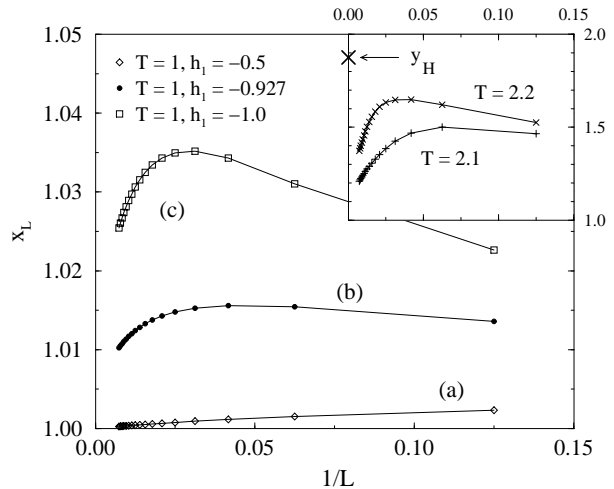,height=7cm}}
\vskip 0.2truecm
\caption{The plots of $x_L$ vs. $1/L$ for different values of temperature and 
surface field where (a) is in the partially wet regime, while (b), (c) and the
curves in the inset are in the completely wet regime. The value of the surface
field for the inset is $h_1 = -1.0$; notice the trend to a crossover towards 
the magnetic exponent $y_{\rm H} = 15/8$ indicated by an arrow in the inset.
}
\label{FIG03}
\end{figure}

Figure~\ref{FIG03} shows a plot of $x_L$ vs. $1/L$ for different values of 
temperatures and surface fields. The set (a) corresponds to the case of partial 
wetting, where the scaling behavior of $x_L$ should be linear in $1/L$ since 
one expects $\gamma = 2 $ and $\alpha = 1$ in Eq. (\ref{complete}).
The fact that these data follow straight lines confirms the behavior predicted 
by the theory.

A clearly different behavior is observed for complete wetting, as shown in the 
sets (b) and (c) and in the inset of Fig.~\ref{FIG03}. In this case $x_L$ has a 
non-monotonic behavior as a function of $1/L$; the numerical data do not follow 
straight lines and seem to approach the value $\alpha = 1$ as $1/L^{\rho}$, 
with $\rho < 1$.
Notice the difference between the data at low temperatures (sets (b) and (c))
and at temperatures not too far from the bulk critical temperature (inset).
In the former case the value of the logarithmic derivative $x_L$ for the largest 
size analyzed is just a few percent off the expected asymptotic value $x_L \to 
\alpha = 1$.
In the latter case $x_L$ is much further away from its expected asymptotic value. 
This is the effect of a crossover towards a scaling of the type (\ref{scalmag}) 
governed by the magnetic exponent $y_{\rm H} = 15/8$, which is shown by a cross
in the inset of Fig.~\ref{FIG03}.

\end{multicols}\widetext
\begin{table}[b]
\caption{The values of the scaling exponent $\alpha$ and the coefficient $A$ 
obtained by the BST extrapolation {\protect \cite{malte}} of the logarithmic 
derivatives $x_L$ in the limit $L \to \infty$, for different values of $T$ 
and $h_1$. 
The third column shows the values of the wetting temperature $T_w (h_1)$, which 
is known exactly {\protect \cite{abrahamlett}} in the two-dimensional Ising model.
In the last column we show the ratio of surface tension $\sigma_0$ and bulk 
magnetization $m_b$ calculated from the exact solution of the two-dimensional
Ising model, which is expected to be identical to the coefficient $A$ (see
Eq. \ref{kelvin}) in the complete wet regime.
The convergence to the expected value $\alpha = 1$ is always very good, but
it is clearly better for partial wetting, as compared to complete wetting.}
\label{T1}
\begin{tabular}{cccccc}
$T$&$h_1$&$T_w (h_1) $&$\alpha$& A & $ \sigma_0/m_b $ \\
\tableline
$1.0$ & $-0.5$ & $1.958\ldots$ & $1.000000(1) $ & $0.99077(1)$ & $1.72891$\\
$1.0$ & $-0.927\ldots$  & $1.0$   & $1.0003(1) $ & $1.729(1)$   & $1.72891$\\
$1.0$ & $-1.0$ &  $0$    & $1.0003(1) $ & $1.729(1)$   & $1.72891$\\
$2.1$ & $-0.5$ & $1.958\ldots$ & $1.004(2) $ & $0.335(2)$   & $0.33508$
\end{tabular}
\end{table}
\begin{multicols}{2}\narrowtext

Table \ref{T1} shows the values of $\alpha$ and $A$ obtained from an accurate 
numerical extrapolation of the DMRG data for different values of the temperature 
and surface field, both at partial ($T < T_w$) and at complete ($T_w \leq T < T_c$)
wetting. For the extrapolation we used the BST algorithm \cite{malte}. Notice the 
excellent agreement with $\alpha = 1$ as predicted by the Kelvin equation (\ref{kelvin}).
Extrapolations are better at low temperatures and at partial wetting, in agreement
with the fact that higher order corrections in $1/L$ vanish more rapidly in the
limit $L \to \infty$ with respect to the complete wetting case.

The last column of Table \ref{T1} shows the ratio between the surface tension 
($\sigma_0$) and bulk spontaneous magnetization ($m_b$), which are known
exactly in the two-dimensional Ising model. According to the Kelvin equation 
(\ref{kelvin}) this ratio should be equal to the coefficient $A$ at complete wetting
where $\cos \theta = 1$. The numerical results agree very well with this; notice also
that at complete wetting the coefficient $A$ does not depend on the surface field
anymore.

\begin{figure}[b]
\centerline{
\psfig{file=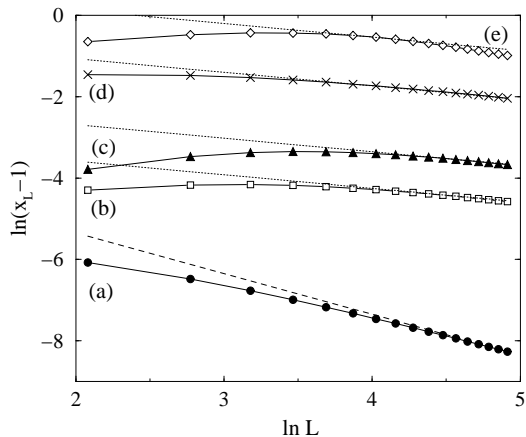,height=6.5cm}}
\vskip 0.2truecm
\caption{
Plots of $\ln ( x_L - 1 )$ vs. $\ln L$ for different values of surface field
and temperature. The different sets correspond to (a) $T = 1.0$, $h_1 = - 0.5$
(b) $T = 1.0$, $h_1 = h_w = - 0.9271$ (c) $T = 1.0$, $h_1 = - 1.0$
(d) $T = 2.1$, $h_1 = - 0.5$ and (e) $T = 2.2$, $h_1 = - 1.0$.
The dashed line has slope $-1$, whereas the dotted lines have slope $-1/3$.
}
\label{FIG04}
\end{figure}

Having found $\alpha = 1$ with high accuracy we can now consider higher order
corrections; setting $\alpha = 1$ in Eq. (\ref{complete}) one obtains:
\begin{eqnarray}
\ln \, ( x_L - 1 ) = \ln \left| \frac B A \right|
- (\gamma - 1) \, \ln L + \ldots
\label{deflogxL}
\end{eqnarray}
Figure~\ref{FIG04} shows a plot of $\ln ( x_L - 1 )$ vs. $\ln L$ for different
temperatures and surface fields. The behavior at partial (a) and complete wetting
(b,c and d) can be clearly distinguished. The straight dashed line in the figure
has slope $-1$ and fits well the asymptotic regime at partial wetting, in good 
agreement with the expected correction-to-scaling exponent $\gamma = 2$.
Other calculations at partial wetting are in good agreement with this 
values as well.

The asymptotic behavior for the sets (b), (c) and (d) of Fig.~\ref{FIG04}
is well-fitted by straight lines in the log-log plot with slope $-1/3$ (dotted 
lines). This is not in agreement with the current theory which predicts a 
correction-to-scaling exponent $\gamma = 5/3$ at complete wetting and one would 
therefore expect an asymptotic slope equal to $1 - \gamma = -2/3$, i.e. twice
as large than that of the dotted lines of Fig.~\ref{FIG04}.
For the set (e) one observes a slight deviation of the data from $-1/3$ towards 
higher slopes at large $L$, but not compatible with $-2/3$. This suggests that 
the large value of $L$ considered here ($L = 144$) is still not sufficient to be 
in the asymptotic regime.
The striking characteristic of the data at complete wetting is that for
a large range of temperatures and surface fields (see sets (b), (c) and (d))
and up to $L = 144$ the correction to scaling seems to be given by a term of
type $1/L^{4/3}$. This behavior will be discussed in more detail in the next 
sections.

\begin{figure}[b]
\centerline{
\psfig{file=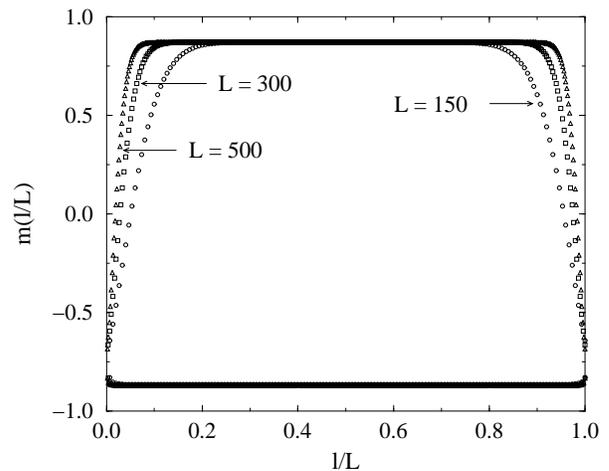,height=7.0cm}}
\vskip 0.2truecm
\caption{The magnetization profiles of the two coexisting phases at $T=2.1$ and 
$h_1=-0.5$, approaching complete wetting, for different strip widths (circles: 
$L=150$, squares: $L=300$ and triangles: $L=500$).}
\label{FIG05}
\end{figure}

To complete the presentation of the DMRG results we show in Fig.~\ref{FIG05} the
magnetization profiles for the two coexisting states. The magnetizations are plotted 
as a function of the scaled variable $l/L$, with $l = 1$, $2$, \ldots, $L$ labeling 
the distance from one of the walls. 
Temperature and surface fields are fixed to $T = 2.1$ and $h_1 = -0.5$ respectively;
this corresponds to a regime of complete wetting ($T_w (h_1 = -0.5)  = 1.958$).
We should also stress that the profiles refer to bulk fields slightly lower and higher
than the pseudo-coexistence field $h_0 (L)$. For bulk field exactly equal to $h_0(L)$ 
the magnetizations of the two coexisting phases are averaged and it is not possible to 
distinguish between them. The positive bulk field favors a bulk phase with positive
magnetization but the negative surface fields favor the adsorption of negative spins
at the boundaries. Since $T \geq  T_w$ the negative spins form a layer that wets the 
walls (as can be seen clearly in the three uppermost profiles of Fig.~\ref{FIG05})
and thickens as $L \to \infty$, or $h \to 0$.
As will be derived in the next section the wetting layer grows for large 
$L$ as $L^{1/3}$, thus it remains macroscopically thin \cite{dietrich}  in the 
thermodynamic limit $L \to \infty$. This can be seen in the figure as well, since in 
the scaled plots of Fig.~\ref{FIG05} the wetting layer flattens towards the wall as 
$L$ increases.

\section{Thermodynamic Theory}
\label{sec:theory}

We briefly review here the theoretical arguments leading to a correction-to-scaling 
exponent $\gamma = 5/3$ for the two-dimensional Ising model, 
following the derivation given by Albano {\it et al.} \cite{albano}. A similar 
derivation, but extended to higher dimensionalities and also to long range 
interaction forces has been presented in Ref. \onlinecite{andy}.

For $T < T_c$ the system can in be in one of the two phases with the
majority of spins pointing either up or down with magnetization profiles
similar to those shown in Fig.~\ref{FIG05}. The total free energy can 
be divided into bulk and surface contributions as follows:
\begin{eqnarray}
f^{\pm} (h, h_1, T) = f_b^{\pm} (h, T) + \frac 2 L
f^{w_\pm}_s (h, h_1, T) 
\label{freen}
\end{eqnarray}
where the superscripts $+$ or $-$ indicate the two phases and the surface term
$f^{w_\pm}$ is the excess free energy for the wall-spin up (or down) phase.
There is no interaction between the surfaces since, as pointed out in the previous 
section, even for $T \geq  T_w$ the surface layers remain very far from each other 
in the limit $L \to \infty$.

Following Ref. \onlinecite{albano} we expand the bulk term to lowest orders in $h$:
\begin{eqnarray}
f_b^{\pm} (h, T) = f_b (T) \mp m_b h + \frac \chi 2 \, h^2 + \ldots
\label{exp1}
\end{eqnarray}
with $m_b > 0 $ the bulk spontaneous magnetization and $\chi$ the 
susceptibility.
For the surface term we write:
\begin{eqnarray}
f_s^{w_\pm} (h, h_1, T) = \sigma^{w_\pm} + m^{\pm}_s (h, h_1, T) \, h
\label{exp2}
\end{eqnarray}
where $\sigma^{w_\pm} = f_s^{w_\pm} (0, h_1, T)$ and $m^{\pm}_s$ is 
the surface excess magnetization.
The condition for phase coexistence is:
\begin{eqnarray}
f^{+} (h, h_1, T) = f^{-} (h, h_1, T)
\end{eqnarray}
From this relation and from Eqs. (\ref{exp1}) and (\ref{exp2}) one finds to 
lowest orders in $1/L$ the following expression for the coexistence bulk field:
\begin{eqnarray}
h_0 (L) = \frac 1 L \,\, \frac{\sigma^{w_+} - \sigma^{w_-}}{m_b - 
\frac 1 L \,\, \left( m_s^+ - m_s^- \right) }
\label{scalh0}
\end{eqnarray}
Using Young's relation \cite{dietrich} $\sigma^{w_+} - \sigma^{w_-} = \sigma_0
\cos \theta$, one finds to leading order in $1/L$ the Kelvin equation
(\ref{kelvin}).

At partial wetting the surface magnetizations $m_s^{\pm}$ remain finite and 
higher order corrections to the Kelvin equation are expected to be of type 
$1/L^2$, in agreement with our numerical results. More interesting is the case 
of complete wetting where the thickness of the two wetting layers in the 
two-dimensional Ising model diverge as $l \sim h^{-1/3}$ for $h \to 0$. From
$h \sim 1/L$ one derives $l \sim L^{1/3}$, which is the result anticipated at 
the end of the previous section. Since the surface magnetization diverges with 
the same power of $h$ as the thickness $l$, i.e. $m_s \sim h^{-1/3} \sim L^{1/3}$, 
from Eq. (\ref{scalh0}) one obtains \cite{albano,andy}:
\begin{eqnarray}
h_0 (L) = \frac A L + \frac B {L^{5/3}} + \ldots
\label{therm}
\end{eqnarray}

\section{Thickness of the wetting layer}
\label{sec:wettinglayer}

\begin{figure}[b]
\centerline{
\psfig{file=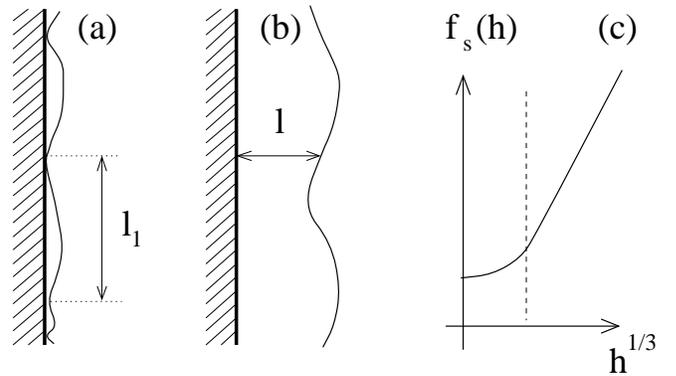,height=5.0cm}}
\vskip 0.2truecm
\caption{
(a,b) The two regimes for approach to complete wetting. In the case (a)
the contacts with the wall produce a surface free energy proportional to
$h^{1/3}$ which results in the correction to the Kelvin equation of the 
type $1/L^{4/3}$. 
In the case (b) when the interface is sufficiently far from the wall this 
free energy is proportional to $h^{2/3}$ and the correction to the Kelvin
equation, according to existing theories \protect\cite{albano,andy},
are expected to be of type $1/L^{5/3}$.
(c) The dependence of the singular part of the surface free energy on the 
field; the vertical dashed line separate the two different scaling regions. 
}
\label{FIG06}
\end{figure}

The argument leading to the expansion (\ref{therm}) is based on the assumption 
that the thickness of the wetting layer $l$ is large enough for the scaling 
relation $m_s \sim l \sim h^{-1/3}$ to be valid. For thin layers one expects 
a deviation from this behavior and consequently from a correction to scaling of
the type $1/L^{5/3}$.
Figures~\ref{FIG06}(a) and \ref{FIG06}(b) show two possible configurations:
A very thin wetting layer with many contacts with the wall (a) and a thicker
one with rare contacts and with an interface well-separated from the wall (b).
Although the DMRG calculations were performed up to large $L$ it seems quite
clear from the magnetization profiles of Fig.~\ref{FIG05} that the typical
configuration of the wetting layer is closer to that of Fig.~\ref{FIG06}(a) 
than to that shown in Fig.~\ref{FIG06}(b).
In the case of a thick wetting layer, as in Fig.~\ref{FIG06}(b), one would 
expect magnetization profiles that start off with a plateau close to the wall
and decay to the bulk magnetization quite far from it. No signs of a plateau
are to be seen in the profiles of Fig.~\ref{FIG05}: The decay to the bulk
magnetization is very rapid and starts already from the boundary spins.

To explain the corrections to the Kelvin equation found in the DMRG calculations 
we have to understand the scaling properties of the singular part of the free 
energy of a thin wetting layer, such as that shown in Fig.~\ref{FIG06}(a).
One expects a free energy proportional to $1/l_1$, i.e. proportional to the
probability of contacts with the wall ($l_1$ is the average distance between 
two contact points).
The scaling behavior of $l_1$ is well known in the literature (see, for instance,
Refs. \onlinecite{abraham,fisher2}) and it is related to the mean first length of a 
return to the ``origin", i.e. the wall, for a random walk which is biased towards the 
wall due to the presence of the bulk field. It has been shown that:
\begin{eqnarray}
l_1 \sim h^{-1/3} .
\label{scall1}
\end{eqnarray}
Assuming thus that the surface free energy (\ref{exp2}) has a singular term 
proportional to $h^{1/3}$ and repeating the analysis of the previous section
one finds a correction-to-scaling term of the type $1/L^{4/3}$.
The case of Fig.~\ref{FIG06}(b) has been treated in the previous section; when $l$,
the thickness of the wetting film, is large enough the singular part of the surface 
free energy is dominated by the magnetic energy of this layer and thus proportional 
to $m_s h \sim l \, h \sim h^{2/3}$.

This simple scaling theory predicts a crossover from thin to thick wetting layer
which results in the different singular behavior of the surface free energy shown 
in Fig.~\ref{FIG06}(c). As a consequence one expects a crossover in the 
correction-to-scaling term from $1/L^{4/3}$ to $1/L^{5/3}$.

\section{The solid-on-solid model}
\label{sec:capill}

In order to obtain a thicker wetting layer and observe the predicted crossover one
could increase the temperature in the DMRG calculations, since entropy favors an
interface far from the wall where more configurations are available. This is indeed
consistent with the behavior of the data of Fig.~\ref{FIG04}; at the largest 
temperature considered (the set (e) corresponds to $T \approx 0.97 \, T_c$) a deviation 
from a correction to scaling of the type $1/L^{4/3}$ towards higher powers of $1/L$
can be clearly observed. This deviation is not yet compatible with the expected 
$1/L^{5/3}$.
However one cannot increase the temperature too much since close to $T_c$ there is
a crossover to a scaling of the type given in Eq. (\ref{scalmag}), which makes the
analysis very difficult.

A second possibility is to enlarge the strip width. Although DMRG calculations for
$L > 144$ are feasible (we have calculated magnetization profiles up to $L = 500$
as shown in Fig.~\ref{FIG05}) it turns out that it is difficult to keep good accuracy
for larger $L$ since $h_0 (L)$ decreases and one needs higher computational effort to
maintain a small relative error on the data.

In this section we analyze the correction to scaling to the Kelvin equation 
(\ref{kelvin}) in the solid-on-solid (SOS) model \cite{abraham} which provides a 
simplified description of the wetting layer. The model does not have a bulk critical 
point (thus no crossover to a scaling of type (\ref{scalmag}) is expected) and it allows 
the study of much larger systems than those available in a DMRG calculation.

In the SOS model a sharp interface is assumed to separate the two phases with all 
spins pointing either up or down; fluctuations within each phase are neglected. 
No overhangs are allowed, so the position of the interface is described by a single 
valued discrete function $l(k)$, where $k$ labels the coordinate position along the 
wall and $l$ is the distance of the interface from the wall. 

Neglecting the interaction between the wetting layers at the two walls we take a 
single wetting layer in a strip of size $L/2$, so the discrete coordinate takes 
the values $l=0$, $1$, \ldots, $L/2$. We consider the one-dimensional transfer matrix, 
whose elements are given by ${\cal T}_{l l^\prime} = {\rm exp}(-\beta  {\cal E}_{l l^\prime})$, 
where $\beta$ is the inverse temperature and the energy is given by \cite{note}:
\begin{eqnarray}
{\cal E}_{l l^\prime} &=& 2 J \, |l - l^\prime| + \left( J + h_1 \right)
\, \left( \theta_l + \theta_{l^\prime}\right) - h_1
\nonumber\\
&+& h \left(l + l^\prime - \frac L 2 \right) ,
\label{energy}
\end{eqnarray}
with:
\begin{eqnarray}
\theta_l = \left\{
\begin{array}{crl}
0 & {\rm if} & l = 0 \\
1 & {\rm if} & l > 0 \\
\end{array}
\right. .
\end{eqnarray}
The configuration $l(k) \equiv 0$ corresponds to all spins up and gives
the ground state energy:
\begin{eqnarray}
\epsilon_0 = -h_1 - \frac{h L}{2} .
\label{groundst}
\end{eqnarray}
The free energy per row is equal to:
\begin{eqnarray}
f_{\rm SOS}^+ (h, h_1, T, L) = - \frac 1 \beta \, \ln \lambda_{\rm max} ,
\label{phasea}
\end{eqnarray}
with $\lambda_{\rm max}$ the largest eigenvalue of the transfer matrix 
$\cal T$.
For the other phase where no wetting layer is present we assume that 
all the spins are pointing down, which gives the free energy:
\begin{eqnarray}
f_{\rm SOS}^- (h, h_1, T, L) = h_1 + \frac{h L}{2} .
\label{phaseb}
\end{eqnarray}
The value of the bulk field which gives phase coexistence, $h_0 (L)$, can be found 
by equating the two free energies in Eqs. (\ref{phasea}) and (\ref{phaseb}), which we
have done numerically for strips of widths up to $L=2000$. The advantage of the SOS
approach is that it allows one to deal with very large systems since the size of the
transfer matrix ${\cal T}$ grows only linearly as function of the strip width $L$.

\begin{figure}[b]
\centerline{
\psfig{file=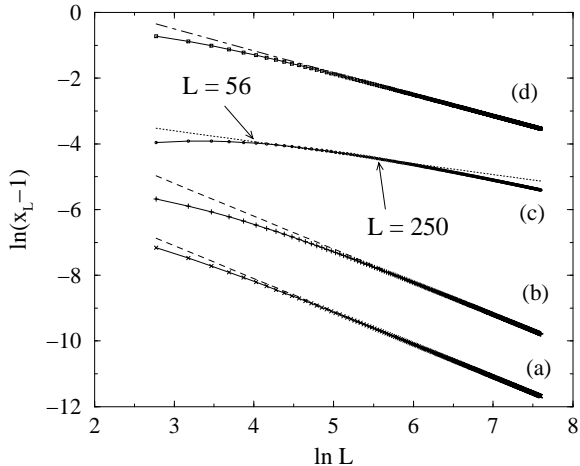,height=7.0cm}}
\vskip 0.2truecm
\caption{
Plot of $\ln (x_L - 1)$ vs. $\ln L$ in the SOS approximation for (a) $T = 1.0$, 
$h_1 = - 0.3$, (b) $T = 1.0$, $h_1 = - 0.7$, (c) $T = 1.0$, $h_1 = - 0.95$ and 
(d) $T = 2.1$, $h_1 = -1.0$.
The set (a) and (b) correspond to the partial wet regime, while (c) and (d) to
complete wetting. The slopes of the straight lines is equal to $-1$ (dashed), 
$-1/3$ (dotted) and $-2/3$ (dashed-dotted).}
\label{FIG07}
\end{figure}

Figure~\ref{FIG07} shows a plot of $\ln (x_L -1)$ vs. $\ln L$, where $x_L$ is given
in Eq. (\ref{defxL}), for various values of temperature and surface field.
The sets (a) and (b) refer to partial wetting and their asymptotic behavior is very 
well fitted by straight lines with slope $-1$, in agreement with a correction to 
scaling of type $1/L^2$ for this case. The other sets refer to $T \geq T_w$, i.e. to 
complete wetting.
The set (c) corresponds to $T=1$ and $h_1 = -0.95$. For a wide range of system sizes 
($56 \leq L \leq 250$) the SOS data follow closely a straight line with slope $-1/3$, 
as also found in the DMRG calculation, but for $L$ large enough the data bends towards 
higher slopes, demonstrating that the correction to scaling found in DMRG calculations 
is not the true asymptotic behavior. The crossover region appears to be quite large since 
still for $L = 2000$ we do not find the expected slope $-2/3$.
At this low temperature $T = 1 \approx 0.4 \, T_c$ the SOS data are also in quantitative 
good agreement with the DMRG results. At higher temperatures this is not the case anymore. 
The set (d) refers to $T = 2.1$, $h_1=-1.0$.
In this case there is a good agreement with the slope $-2/3$ of the dashed-dotted line, 
as expected on the basis of the scaling theory of Albano {\it et al.} \cite{albano}
and of Parry and Evans \cite{andy}. It is also important to point out that for the
full Ising calculation at the same temperature and surface field we still find a
slope $-1/3$, which suggests that the Ising interface is closer to the wall than
the SOS interface for the same values of surface field and temperature.

We have also calculated $\langle l \rangle$, the average thickness of the wetting 
layer, from the SOS Hamiltonian. For the sets (c) and (d) and $L = 200$ we found 
$\langle l \rangle = 1.3 $ and  $\langle l \rangle = 11.5$ respectively (these 
values are expressed in terms of lattice units). 
Notice that for a system of size $L = 200$ one observes corrections to scaling of
the type $1/L^{4/3}$ and $1/L^{5/3}$ in the sets (c) and (d) respectively and the 
large differencee in the values of $\langle l \rangle$ is consistent with a crossover 
from a thin to a thick wetting layer discussed in the previous section.

\section{Conclusions}
\label{sec:concl}

In this paper we have analyzed the influence of wetting on the thermodynamics of
an Ising model in a confined geometry with equal surface fields at the boundaries.
Firstly, we verified with high numerical accuracy that the coexistence bulk 
field scales as $1/L$ ($L$ is the distance between the boundaries) in agreement with 
the predictions of the Kelvin equation. Then our study focused on the higher order 
corrections to this equation which are expected to be analytic (non-analytic) at 
partial (complete) wetting.

Using density-matrix renormalization techniques we found that for a large range
of surface fields and temperature higher order corrections are not compatible with
$1/L^{5/3}$, which is the behavior predicted by the existing theory, but they are of 
type $1/L^{4/3}$. 
We have shown that this apparent disagreement is due to the fact that even for the 
large sizes considered ($L = 144$) the wetting layer has a limited thickness, so that 
the singular part of the surface free energy which determines the 
correction-to-scaling behavior is dominated by the contacts with the walls.
In this case a simple random walk argument indeed predicts a correction term of 
the type $1/L^{4/3}$ for a thin wetting layer.
We have analyzed the problem also using a solid-on-solid model, which provides a
simplified description of the wetting layer, but allows the study of much larger
systems (strips of widths up to $L = 2000$ were considered). The results of this
analysis are consistent with the $1/L^{4/3}$ correction for small thicknesses of
the wetting layer, but also show a crossover to a correction of type $1/L^{5/3}$
for sufficiently thick layers.

{\bf Acknowledgements} -
We thank A. O. Parry for stimulating discussions that lead to the present study.
Discussions with R.~Blossey, C.~J.~Boulter, J.~O.~Indekeu and E.~Montevecchi are 
also gratefully acknowledged. A.~D. would like to thank the Institute for Theoretical
Physics of the K. U. Leuven, where part of this work was done, for kind hospitality.
E.~C. was financially supported by the Fund for Scientific Research of Flanders
(FWO G.0239.96).

\end{multicols}

\begin{references}
\bibitem{note} For convenience we have set to zero the ground state energy of the
Hamiltonian (\ref{ham}) for $h_1 = h = 0$, by a global shift of the energy scale.
\end{references}
\end{document}